# A Matrix Decomposition Method for Odd-Type Gaussian Normal Basis Multiplication

Kittiphon Phalakarn and Athasit Surarerks
Department of Computer Engineering, Faculty of Engineering
Chulalongkorn University
Bangkok, Thailand
e-mail: kittiphon.phalakarn@gmail.com, athasit.s@chula.ac.th

*Abstract*—**Normal basis is used in many applications because of the efficiency of the implementation. However, most space complexity reduction techniques for binary field multiplier are applicable for only optimal normal basis or Gaussian normal basis of even type. There are 187 binary fields GF($2^k$) for $k$ from 2 to 1,000 that use odd-type Gaussian normal basis. This paper presents a method to reduce the space complexity of odd-type Gaussian normal basis multipliers over binary field GF($2^k$). The idea is adapted from the matrix decomposition method for optimal normal basis. The result shows that our space complexity reduction method can reduce the number of XOR gates used in the implementation comparing to previous works with a small trade-off in critical path delay.**

## I. Introduction

The binary fields GF($2^k$) are widely used and have many applications in cryptography and coding theory. The main operations of the fields include addition, multiplication, and squaring. The field elements are usually represented using polynomial basis or normal basis. Between these two bases, normal basis is considered as a better choice for hardware implementation because squaring can be implemented efficiently by a bit cyclic shift which is considered as a free operation. However, normal basis multipliers are complexed and space-consuming. The first implementation of normal basis multiplier was presented by Massey and Omura in [1], and many techniques to reduce space complexity of normal basis multipliers were developed since then. To have smaller space complexity, the special classes of normal basis called an optimal normal basis (ONB) and a Gaussian normal basis (GNB) are used as a basis of the binary field. The space complexity, illustrated as a number of gates used in the implementation, is bounded when using these bases.

To reduce the space complexity further, several methods were proposed. In this paper, we are interested in reduction techniques for bit-parallel multipliers which produce a result in one clock cycle. Examples of such techniques use new formulation [2,3,4,5], common subexpression algorithms [6], and a matrix decomposition method [7]. However, most techniques, including those mentioned above, are applicable for only optimal normal basis and Gaussian normal basis of even type. From our best knowledge, we do not find any work in literature try to reduce space complexity of binary field multipliers that use odd-type Gaussian normal basis. Nevertheless, from IEEE standard [8], there are 187 binary fields GF($2^k$) for $k$ from 2 to 1,000 that use odd-type Gaussian normal basis. Some first few GF($2^k$) that use odd-type Gaussian normal basis are $k$ = 20 (type 3), 22 (type 3), 34 (type 9), 42 (type 5), 44 (type 9), 46 (type 3), and 54 (type 3).

That motivates us to consider a space complexity reduction technique for binary field multipliers that use odd-type Gaussian normal basis. This paper adapts an idea of the matrix decomposition method for optimal normal basis in [7] with odd-type Gaussian normal basis. We first prove a property of odd-type Gaussian normal basis and then introduce the decomposition method for reducing space complexity of the multipliers in this case. The result shows that our method can reduce the number of XOR gates used in the implementation comparing to previous works with a small trade-off in an increase of critical path delay.

## II. Preliminaries

In this section, we review some backgrounds on normal basis, optimal normal basis, Gaussian normal basis, and the matrix decomposition method for optimal normal basis multiplication.

### A. Normal Basis

Normal basis is one way of representing elements in the binary field GF($2^k$). The definition is as follows.

**Definition 1 (Normal Basis).** *Given a binary field* GF($2^k$), *a basis* <$\beta, \beta^2, \beta^4, ..., \beta^{\exp(2,k-1)}$> *with* $\beta \in$ GF($2^k$) *is called a* normal basis *if every* $\alpha \in$ GF($2^k$) *can be uniquely written as*

$$\alpha = \alpha_0\beta + \alpha_1\beta^2 + \alpha_2\beta^4 + ... + \alpha_{k-1}\beta^{\exp(2,k-1)} \; ; \; \alpha_i \in \{0,1\}.$$

*In this paper, we use* $\beta_i = \beta^{\exp(2,i)}$ *to simplify the notation.*

When using normal basis, the element addition is bitwise XOR, and squaring, $\alpha' = \alpha^2$, is a bit cyclic shift ($\alpha'_i = \alpha_{i-1}$ for $1 \leq i < k$ and $\alpha'_0 = \alpha_{k-1}$) since $\beta_k = \beta_0$. For multiplication, we explain the method in Example 1.

**Example 1.** *Consider* $GF(2^3) = \{0, 1, x, x+1, x^2, x^2+1, x^2+x, x^2+x+1\}$ *with an irreducible polynomial* $P(x) = x^3+x+1$ *and normal basis* $<\beta_0 = x+1, \beta_1 = x^2+1, \beta_2 = x^2+x+1>$. *To find* $ab = (a_0\beta_0+a_1\beta_1+a_2\beta_2)(b_0\beta_0+b_1\beta_1+b_2\beta_2) = c_0\beta_0+c_1\beta_1+c_2\beta_2$, *we multiply each term together as defined by Table I.*

The multiplication result is the summation of all cells in the table. Here, $\beta^3$, $\beta^5$, and $\beta^6$ are not expressed as a summation of the basis. They can be expressed as in Table II. Hence, each bit of the result $c$ is as follows;

$$c_0 = a_0b_1+a_1b_0+a_1b_2+a_2b_1+a_2b_2$$
$$c_1 = a_0b_0+a_0b_2+a_1b_2+a_2b_0+a_2b_1 \quad (1)$$
$$c_2 = a_0b_1+a_0b_2+a_1b_0+a_1b_1+a_2b_0.$$

These can be expressed as multiplication matrices in Table III. We denote the number of 1's in each multiplication matrix as $C_N$ (all matrices have the same number of 1's). The number of AND gates used is $k^2$, and, from (1), the number of XOR gates used is $k(C_N-1)$. In this example, $C_N$ is 5, and the number of XOR gates used is 12. □

### B. Optimal Normal Basis (ONB)

From the previous subsection, we can see that $C_N$ affects the space complexity. This subsection provides a detail about optimal normal basis (ONB) which gives the smallest possible $C_N = 2k-1$. This is discussed in [7,9].

**Theorem 1 (ONB).** *For* $GF(2^k)$, *the ONB exists in one of the following cases.*
- *Type 1: If $k+1$ is a prime and 2 is primitive in $GF(k+1)^*$, then every primitive $(k+1)$th root of unity in $GF(2^k)$ generates an ONB for $GF(2^k)$.*
- *Type 2: If $2k+1$ is a prime and either i) 2 is primitive in $GF(2k+1)^*$, or ii) $2k+1 \equiv 3 \pmod{4}$ and 2 generates quadratic residues in $GF(2k+1)^*$, then $\beta = \gamma + \gamma^{-1}$ generates an ONB for $GF(2^k)$, where $\gamma$ is any primitive $(2k+1)$th root of unity in $GF(2^{2k})$.*

*The number of 1's in the multiplication matrix when using ONB is equal to $2k-1$ which is the smallest possible number of 1's in the multiplication matrix.*

TABLE I. MULTIPLICATION OF $GF(2^3)$

|  | $b_0\beta_0$ | $b_1\beta_1$ | $b_2\beta_2$ |
|---|---|---|---|
| $a_0\beta_0$ | $a_0b_0\ \beta^2$ | $a_0b_1\ \beta^3$ | $a_0b_2\ \beta^5$ |
| $a_1\beta_1$ | $a_1b_0\ \beta^3$ | $a_1b_1\ \beta^4$ | $a_1b_2\ \beta^6$ |
| $a_2\beta_2$ | $a_2b_0\ \beta^5$ | $a_2b_1\ \beta^6$ | $a_2b_2\ \beta^8$ |

TABLE II. MULTIPLICATION OF $GF(2^3)$ IN NORMAL BASIS

|  | $b_0\beta_0$ | $b_1\beta_1$ | $b_2\beta_2$ |
|---|---|---|---|
| $a_0\beta_0$ | $a_0b_0\ \beta^2$ | $a_0b_1\ (\beta+\beta^4)$ | $a_0b_2\ (\beta^2+\beta^4)$ |
| $a_1\beta_1$ | $a_1b_0\ (\beta+\beta^4)$ | $a_1b_1\ \beta^4$ | $a_1b_2\ (\beta+\beta^2)$ |
| $a_2\beta_2$ | $a_2b_0\ (\beta^2+\beta^4)$ | $a_2b_1\ (\beta+\beta^2)$ | $a_2b_2\ \beta$ |

TABLE III. MULTIPLICATION MATRICES OF $GF(2^3)$

| $c_0$ |  |  |
|---|---|---|
| 0 | 1 | 0 |
| 1 | 0 | 1 |
| 0 | 1 | 1 |

| $c_1$ |  |  |
|---|---|---|
| 1 | 0 | 1 |
| 0 | 0 | 1 |
| 1 | 1 | 0 |

| $c_2$ |  |  |
|---|---|---|
| 0 | 1 | 1 |
| 1 | 1 | 0 |
| 1 | 0 | 0 |

### C. Gaussian Normal Basis (GNB)

The limitation of optimal normal basis is that they do not exist for some $k$. For a binary field $GF(2^k)$ without optimal normal basis, some nearly optimal normal bases with small $C_N$ are proposed in [10,11] called Gaussian normal basis (GNB).

**Theorem 2 (GNB of type $T$).** *For any positive integer $T$ and $GF(2^k)$, the GNB of type $T$ exists if two following conditions are satisfied.*
- *$Tk+1$ is a prime.*
- *$s$ is the order of 2 in $GF(Tk+1)^*$ and $\gcd(Tk/s,k) = 1$.*

*Let $\gamma$ be any primitive $(Tk+1)$th root of unity in $GF(2^{Tk})$, and $\lambda$ be any primitive $T$th root of unity in $GF(Tk+1)$, then*

$$\beta = \Sigma_{i=0\ to\ T-1}\ \exp(\gamma,\lambda^i)$$

*generates a Gaussian normal basis of type $T$ for $GF(2^k)$ with the number of 1's in the multiplication matrix satisfies*

$$C_N \leq Tk-1;\quad \text{if } T \text{ is even,}$$
$$C_N \leq (T+1)k-T;\quad \text{if } T \text{ is odd.}$$

We can see that ONB is considered as type 1 GNB and type 2 GNB. Since $C_N$ depends on the value of $T$, small $T$ that satisfies the conditions is typically used for each $k$.

**Example 2.** *Consider GNB of type $T = 3$ for $GF(2^4)$. Both $T$ and $k$ satisfy the conditions. It is obtained that $\lambda \in \{3,9\}$.*

Let $\gamma$ be any primitive 13th root of unity in $GF(2^{12})$, then $\beta = \gamma+\gamma^3+\gamma^9$ (for $\lambda=3$) or $\beta = \gamma+\gamma^9+\gamma^{81} = \gamma+\gamma^3+\gamma^9$ (for $\lambda=9$). Hence, $\beta$, $\beta^2 = \gamma^2+\gamma^5+\gamma^6$, $\beta^4 = \gamma^4+\gamma^{10}+\gamma^{12}$, and $\beta^8 = \gamma^7+\gamma^8+\gamma^{11}$ form a type 3 GNB. All multiplication matrices using this basis are shown in Table IV.

Here, $C_N = 9$ is smaller than $(T+1)k-T = 13$ but greater than $2k-1 = 7$. By the fact that $\gamma$ is a primitive 13th root of unity ($\gamma \neq 1$), the basis can be expressed in polynomial. Thus, $(\gamma^{13}-1)/(\gamma-1) = 0$, and $1+\gamma+\gamma^2+...+\gamma^{12} = 0 = 1+\beta+\beta^2+\beta^4+\beta^8$. Since $\beta^3 = \beta+\beta^2+\beta^8$, therefore $1+\beta^3+\beta^4 = 0$. The $\beta$ that satisfies $1+\beta^3+\beta^4 = 0$ depends on the irreducible polynomial $P(x)$. For example, if $P(x) = x^4+x^3+x^2+x+1$, one possible $\beta$ is $x+1$ which generates a basis $<x+1, x^2+1, x^3+x^2+x, x^3+1>$ that is type 3 GNB for $GF(2^4)$. This example also shows a property of GNB that $\beta_0+\beta_1+\beta_2+...+\beta_{k-1} = 1$. □

### D. Matrix Decomposition Method for ONB Multiplication

The matrix decomposition method [7] is used for reducing the number of XOR gates. This technique can be applied with multipliers of binary field $GF(2^k)$ with ONB.

TABLE IV. MULTIPLICATION MATRICES OF TYPE 3 GNB FOR $GF(2^4)$

| $c_0$ |  |  |  |
|---|---|---|---|
| 0 | 1 | 1 | 1 |
| 1 | 0 | 1 | 0 |
| 1 | 1 | 0 | 0 |
| 1 | 0 | 0 | 1 |

| $c_1$ |  |  |  |
|---|---|---|---|
| 1 | 1 | 0 | 0 |
| 1 | 0 | 1 | 1 |
| 0 | 1 | 0 | 1 |
| 0 | 1 | 1 | 0 |

| $c_2$ |  |  |  |
|---|---|---|---|
| 0 | 0 | 1 | 1 |
| 0 | 1 | 1 | 0 |
| 1 | 1 | 0 | 1 |
| 1 | 0 | 1 | 0 |

| $c_3$ |  |  |  |
|---|---|---|---|
| 0 | 1 | 0 | 1 |
| 1 | 0 | 0 | 1 |
| 0 | 0 | 1 | 1 |
| 1 | 1 | 1 | 0 |

For type 1 ONB, it is proved that $a_i b_{(i+k/2) \bmod k}$, for all $0 \leq i < k$, is used in every bit of the result, and other $a_i b_j$ is used once in exactly one bit of the result. We can first calculate the summation of the common terms in order to reduce the number of XOR gates from $k(2k-2)$ to $k^2-1$. The critical path delay is $T_A + [1+\log_2(k)]T_X$ where $T_A$ is a delay of a two-input AND gate and $T_X$ is a delay of a two-input XOR gate.

**Example 3.** *Table V shows multiplication matrices of type 1 ONB for* $GF(2^4)$. *Without using the decomposition method, 24 XOR gates are used (6 XOR gates for each bit of c). By applying the method, each bit can be calculated as follows;*

$$\omega = a_0 b_2 + a_1 b_3 + a_2 b_0 + a_3 b_1$$
$$c_0 = \omega + a_1 b_2 + a_2 b_1 + a_3 b_3$$
$$c_1 = \omega + a_0 b_0 + a_2 b_3 + a_3 b_2$$
$$c_2 = \omega + a_0 b_3 + a_1 b_1 + a_3 b_0$$
$$c_3 = \omega + a_0 b_1 + a_1 b_0 + a_2 b_2.$$

The implementation uses 15 XOR gates in total. □

For type 2 ONB, it is proved that each summation $a_i b_j + a_j b_i$, for all $0 \leq i < j < k$, is used twice in the equation, and $a_i b_i$, for all $0 \leq i < k$, is used once. We can first calculate every summation $a_i b_j + a_j b_i$, and use them for reducing the number of XOR gates from $k(2k-2) = 2k(k-1)$ to $1.5k(k-1)$. The critical path delay is $T_A + [1+\log_2(k)]T_X$.

**Example 4.** *The basis used in Example 1 is type 2 ONB. By applying the method, we rewrite (1) as follows;*

$$\mu_{01} = a_0 b_1 + a_1 b_0$$
$$\mu_{02} = a_0 b_2 + a_2 b_0$$
$$\mu_{12} = a_1 b_2 + a_2 b_1$$
$$c_0 = \mu_{01} + \mu_{12} + a_2 b_2$$
$$c_1 = \mu_{02} + \mu_{12} + a_0 b_0$$
$$c_2 = \mu_{01} + \mu_{02} + a_1 b_1.$$

Here, the number of XOR gates is reduced from 12 to 9. □

### III. PROPERTY OF ODD-TYPE GAUSSIAN NORMAL BASIS

Most space complexity reduction methods are proposed for optimal normal basis or even-type Gaussian normal basis multipliers since the properties of these bases are proved to be useful for implementation. For ONB, the structure of multiplication matrices is proved in [7], and for even-type GNB with odd $k$, the property that row $k-i$, $1 \leq i \leq (k-1)/2$, is the $i$-fold cyclic shift of row $i$ is proved in [3]. In this section, we study a property of odd-type GNB.

TABLE V. MULTIPLICATION MATRICES OF TYPE 1 ONB FOR $GF(2^4)$

| $c_0$ | | | | $c_1$ | | | | $c_2$ | | | | $c_3$ | | | |
|---|---|---|---|---|---|---|---|---|---|---|---|---|---|---|---|
| 0 | 0 | 1 | 0 | 1 | 0 | 1 | 0 | 0 | 0 | 1 | 1 | 0 | 1 | 1 | 0 |
| 0 | 0 | 1 | 1 | 0 | 0 | 0 | 1 | 0 | 1 | 0 | 1 | 1 | 0 | 0 | 1 |
| 1 | 1 | 0 | 0 | 1 | 0 | 0 | 1 | 1 | 0 | 0 | 0 | 1 | 0 | 1 | 0 |
| 0 | 1 | 0 | 1 | 0 | 1 | 1 | 0 | 1 | 1 | 0 | 0 | 0 | 1 | 0 | 0 |

In our proofs for odd-type $T$ GNB for $GF(2^k)$, by Theorem 2, we assume the followings.
- $T$ is odd and $Tk+1$ is a prime, hence $k$ is even.
- $s$ is the order of 2 in $GF(Tk+1)^*$ and $\gcd(Tk/s,k) = 1$.
- $\gamma$ is any primitive $(Tk+1)$th root of unity in $GF(2^{Tk})$.
- $\lambda$ is any primitive $T$th root of unity in $GF(Tk+1)$.
- $\beta = \Sigma_{i=0 \text{ to } T-1} \exp(\gamma,\lambda^i)$ generates GNB, and $\beta_i = \exp(\gamma,2^i)+\exp(\gamma,2^i\lambda)+\exp(\gamma,2^i\lambda^2)+...+\exp(\gamma,2^i\lambda^{T-1})$.

**Lemma 1.** *If $\lambda$ is a primitive $T$th root of unity in* $GF(Tk+1)$, *then $1, \lambda, \lambda^2, \lambda^3, ..., \lambda^{T-1}$ are all unique $T$th roots of unity in* $GF(Tk+1)$.

*Proof.* From the definition of primitive $T$th root of unity, $\lambda \neq 1$, $\lambda^T = 1$, and $\lambda^i \neq 1$ for all $1 \leq i < T$. If $\lambda^i = \lambda^j$ for some $1 \leq i < j < T$, we have that $\lambda^{j-i} = 1$ which is impossible. Hence, it is obtained that $\lambda^i \neq \lambda^j$. Because, for $0 \leq i < T$, $(\lambda^i)^T = (\lambda^T)^i = 1$, therefore $1, \lambda, \lambda^2, \lambda^3, ..., \lambda^{T-1}$ are all unique $T$th roots of unity in $GF(Tk+1)$. ∎

Next, we study a property of multiplication matrices of odd-type GNB.

**Lemma 2.** *For odd-type $T$ GNB for* $GF(2^k)$, *the pair $a_i b_{(i+k/2) \bmod k}$, $0 \leq i < k$, is used in $k-T+1$ bits of the result.*

*Proof.* The pair $a_i b_{(i+k/2) \bmod k}$ is used in the case of the multiplication of $\beta_i \beta_{(i+k/2) \bmod k} = \exp(\beta_0 \beta_{k/2}, 2^i)$. Since squaring is a bit cyclic shift, only $\beta_0 \beta_{k/2}$ is considered.

From Theorem 2, $\beta_0 = \gamma + \exp(\gamma,\lambda) + \exp(\gamma,\lambda^2) + ... + \exp(\gamma,\lambda^{T-1})$ and $\beta_{k/2} = \exp(\beta_0, 2^{k/2}) = \exp(\gamma,2^{k/2}) + \exp(\gamma,2^{k/2}\lambda) + \exp(\gamma,2^{k/2}\lambda^2) + ... + \exp(\gamma,2^{k/2}\lambda^{T-1})$. We try to express $2^{k/2}$ in term of $\lambda$.

Since $\gcd(Tk/s,k) = 1$, it can be determined that $k \mid s$ and $s \mid Tk$. Let $s = pk$ with $p \mid T$. From the definition of $s$ as the order of 2 in $GF(Tk+1)^*$, then $2^s = 2^{pk} = 1$. Thus, $(2^{k/2})^p = -1$. As $p \mid T$, the $p$th root of unity is one of the $T$th root of unity. Hence, $2^{k/2}$ must be in the set $\{-1, -\lambda, -\lambda^2, -\lambda^3, ..., -\lambda^{T-1}\}$ by Lemma 1.

Suppose that $2^{k/2} = -\lambda^i$ for some $0 \leq i < T$. We can rewrite $2^{k/2}$ using $\lambda$ for $\beta_{k/2}$ as $\exp(\gamma,-\lambda^i) + \exp(\gamma,-\lambda^{i+1}) + \exp(\gamma,-\lambda^{i+2}) + ... + \exp(\gamma,-\lambda^{i+T-1})$. The product of $\beta_0$ and $\beta_{k/2}$ can be formed into $T$ groups as follows;

$$\beta_0 \beta_{k/2} = \Gamma_0 + \Gamma_1 + ... + \Gamma_{T-1}$$
$$\Gamma_0 = \exp(\gamma, 1-\lambda^i) + \exp(\gamma, \lambda-\lambda^{i+1}) + ... + \exp(\gamma, \lambda^{T-1}-\lambda^{i+T-1})$$
$$\Gamma_1 = \exp(\gamma, 1-\lambda^{i+1}) + \exp(\gamma, \lambda-\lambda^{i+2}) + ... + \exp(\gamma, \lambda^{T-1}-\lambda^i)$$
$$...$$
$$\Gamma_{T-1} = \exp(\gamma, 1-\lambda^{i+T-1}) + \exp(\gamma, \lambda-\lambda^i) + ... + \exp(\gamma, \lambda^{T-1}-\lambda^{i+T-2}).$$

Let $j = T-i \bmod T$. It is obtained that $\lambda^{i+j} = 1$ which makes $\Gamma_j = 1+1+...+1$ (with $T$ 1's) = 1. Since other $\Gamma_l$ is equal to some unique $\beta_{l'}$, therefore $\beta_0 \beta_{k/2}$ is a summation of 1 and $T-1$ basis elements. From the property that $\beta_0 + \beta_1 + ... + \beta_{k-1} = 1$, $\beta_0 \beta_{k/2}$ is then a summation of $k-(T-1)$, which is equal to $k-T+1$, basis elements, and $a_i b_{(i+k/2) \bmod k}$ is used in exactly $k-T+1$ bits of the result $c$. ∎

**Example 5.** *From Example 2, let $\lambda = 3$. It is obtained that $2^{k/2} = 4 = -9$. Thus, $\beta_0 = \gamma + \gamma^3 + \gamma^9$ and $\beta_2 = \gamma^{-9} + \gamma^{-1} + \gamma^{-3}$. Then, $\beta_0\beta_2$ can be written as $\Gamma_0 + \Gamma_1 + \Gamma_2$ where $\Gamma_0 = \gamma^{-8} + \gamma^2 + \gamma^6 = \beta_1$, $\Gamma_1 = 1+1+1 = 1$, and $\Gamma_2 = \gamma^{-2} + \gamma^{-6} + \gamma^8 = \beta_3$. Therefore, $\beta_0\beta_2 = \beta_1 + 1 + \beta_3 = \beta_0 + \beta_2$, and $a_0b_2$ is used in $k-T+1 = 2$ bits of $c$, namely $c_0$ and $c_2$. (The case where $\lambda = 9$ is similar.)*

Consider type 3 GNB for $GF(2^6)$ as another example. The multiplication matrices are shown in Table VI. Each $a_ib_{(i+k/2) \bmod k}$, $0 \leq i < k$, is used in $k-T+1 = 4$ bits of $c$. □

From Lemma 2 and the bit cyclic shifting property of squaring, in a multiplication matrix of each bit of $c$, exactly $k-T+1$ from $k$ cells of $a_ib_{(i+k/2) \bmod k}$, $0 \leq i < k$, are 1's and the remaining $T-1$ cells are 0's. Note that all the multiplication matrices are symmetric, and there is only one 1 in the top-left to bottom-right diagonal of each multiplication matrix.

### IV. MATRIX DECOMPOSITION METHOD FOR ODD-TYPE GAUSSIAN NORMAL BASIS MULTIPLICATION

We use the property proved in the previous section to reduce the space complexity of the implementation of the multipliers for odd-type GNB for $GF(2^k)$.

Since $k > T$ in most cases, $k-T+1$ cells of $a_ib_{(i+k/2) \bmod k}$, $0 \leq i < k$, are 1's, and $T-1$ cells are 0's, we can reduce the redundancy by adding all $a_ib_{(i+k/2) \bmod k}$ in order to calculate each bit of the result. Our proposed decomposition method is a combination of the decomposition method for type 1 and type 2 ONB. The detail of our method is as follows.

*Step 1:* Construct the multiplication matrices of odd-type GNB for $GF(2^k)$.

*Step 2:* Calculate every multiplication $a_ib_j$ for all $0 \leq i, j < k$. This step uses $k^2$ AND gates.

*Step 3:* Calculate $\mu_{ij} = a_ib_j + a_jb_i$ for all $0 \leq i < j < k$. This step uses $k(k-1)/2$ XOR gates.

*Step 4:* Calculate $\omega = \Sigma_{i=0 \text{ to } k/2-1} \mu_{i,(i+k/2) \bmod k}$. This step uses $k/2-1$ XOR gates.

*Step 5:* For each bit $c_i$, $0 \leq i < k$, the result comes from three parts: $a_{(i-1 \bmod k)}b_{(i-1 \bmod k)}$, $\omega$, and all necessary $\mu_{ij}$. The number of $\mu_{ij}$ used in this step is $(C_N-k+2T-3)/2$. Hence, this step uses $(C_N-k+2T-3)/2+1$ XOR gates for each bit of the result. The total number of XOR gates used for every bit of the result in this step is $k(C_N-k+2T-3)/2+k$.

The total number of AND gates used in the implementation is $k^2$ and the total number of XOR gates is $k(k-1)/2 + k/2-1 + k(C_N-k+2T-3)/2+k = (k/2)(C_N+2T-1)-1$. See Example 6 for more understanding.

**Example 6.** *The bit $c_0$ and $c_1$ in Table VI are decomposed as shown in Table VII. Table VII (b) and (e) represent $\omega$. We show how to calculate $c_0$ and $c_1$ from each part mentioned in Step 5 as follows;*

$$\omega = \mu_{03} + \mu_{14} + \mu_{25}$$
$$c_0 = a_5b_5 + \omega + (\mu_{02} + \mu_{05} + \mu_{12} + \mu_{13} + \mu_{14} + \mu_{24} + \mu_{45})$$
$$c_1 = a_0b_0 + \omega + (\mu_{01} + \mu_{05} + \mu_{13} + \mu_{23} + \mu_{24} + \mu_{25} + \mu_{35}).$$

By our decomposition method, the number of XOR gates used is reduced from 96 to 65. □

To calculate the critical path delay of our implementation, we propose the design as illustrated in Fig. 1. In order to have a small delay, all summations are performed as a binary tree, i.e. using $(\log_2 n)T_X$ time units to sum up $n$ elements. From our design, the critical path delay is $T_A + [1+\log_2(C_N-k+2T-1)]T_X$.

**Theorem 3.** *Our matrix decomposition method for odd-type Gaussian normal basis multiplication in $GF(2^k)$ computes the result using $k^2$ AND gates, $(k/2)(C_N+2T-1)-1$ XOR gates, and a total delay of $T_A + [1+\log_2(C_N-k+2T-1)]T_X$.*

*Proof.* By Lemma 2 and the design in Fig. 1. ∎

### V. DISCUSSION

In this section, we compare our result with previous methods. The first method is the straightforward method described in Example 1. The second and third methods from [12] are the best bit-parallel multipliers which are applicable for all normal bases, also odd-type GNB, we have found. The number of AND gates, XOR gates, and the critical path delay of each method are compared in Table VIII.

We can see that our method reduces the number of XOR gates from previous methods. The number of XOR gates in our method depends on $T$ which is less than $k$. The critical path delay of our method is close to others as by one $T_X$.

TABLE VI. MULTIPLICATION MATRICES OF TYPE 3 GNB FOR $GF(2^6)$

| $c_0$ | | | | | |
|---|---|---|---|---|---|
| 0 | 0 | 1 | 1 | 0 | 1 |
| 0 | 0 | 1 | 1 | 0 | 0 |
| 1 | 1 | 0 | 0 | 1 | 1 |
| 1 | 1 | 0 | 0 | 0 | 0 |
| 0 | 0 | 1 | 0 | 0 | 1 |
| 1 | 0 | 1 | 0 | 1 | 1 |

| $c_1$ | | | | | |
|---|---|---|---|---|---|
| 1 | 1 | 0 | 1 | 0 | 1 |
| 1 | 0 | 0 | 1 | 1 | 0 |
| 0 | 0 | 0 | 1 | 1 | 0 |
| 1 | 1 | 1 | 0 | 0 | 1 |
| 0 | 1 | 1 | 0 | 0 | 0 |
| 1 | 0 | 0 | 1 | 0 | 0 |

| $c_2$ | | | | | |
|---|---|---|---|---|---|
| 0 | 1 | 0 | 0 | 1 | 0 |
| 1 | 1 | 1 | 0 | 1 | 0 |
| 0 | 1 | 0 | 0 | 1 | 1 |
| 0 | 0 | 0 | 0 | 1 | 1 |
| 1 | 1 | 1 | 1 | 0 | 0 |
| 0 | 0 | 1 | 1 | 0 | 0 |

| $c_3$ | | | | | |
|---|---|---|---|---|---|
| 0 | 0 | 0 | 1 | 1 | 0 |
| 0 | 0 | 1 | 0 | 0 | 1 |
| 0 | 1 | 1 | 1 | 0 | 1 |
| 1 | 0 | 1 | 0 | 0 | 1 |
| 1 | 0 | 0 | 0 | 0 | 1 |
| 0 | 1 | 1 | 1 | 1 | 0 |

| $c_4$ | | | | | |
|---|---|---|---|---|---|
| 0 | 0 | 1 | 1 | 1 | 1 |
| 0 | 0 | 0 | 0 | 1 | 1 |
| 1 | 0 | 0 | 1 | 0 | 0 |
| 1 | 0 | 1 | 1 | 1 | 1 |
| 1 | 1 | 0 | 1 | 0 | 1 |
| 1 | 1 | 0 | 0 | 0 | 0 |

| $c_5$ | | | | | |
|---|---|---|---|---|---|
| 0 | 1 | 1 | 0 | 0 | 0 |
| 1 | 0 | 0 | 1 | 1 | 1 |
| 1 | 0 | 0 | 0 | 0 | 1 |
| 0 | 1 | 0 | 0 | 1 | 1 |
| 0 | 1 | 0 | 1 | 1 | 1 |
| 0 | 1 | 1 | 0 | 1 | 0 |

TABLE VII. DECOMPOSITION OF $C_0$ AND $C_1$ OF TYPE 3 GNB FOR $GF(2^6)$

(a), (b), (c), (d), (e), (f) — matrix decomposition tables.

TABLE VIII. COMPARISON OF BIT-PARALLEL
ODD-TYPE GAUSSIAN NORMAL BASIS MULTIPLIERS

| Method | #AND | #XOR | Critical Path Delay |
|---|---|---|---|
| naive | $k^2$ | $k(C_N-1)$ | $T_A + [\log_2(C_N)]T_X$ |
| XEBP [12] | $k^2$ | $(k/2)(C_N+k-2)$ | $T_A + [\log_2(C_N)]T_X$ |
| AEBP [12] | $(k/2)(k-1)$ | $(k/2)(C_N+2k-3)$ | $T_A + [\log_2(C_N)]T_X$ |
| our method | $k^2$ | $(k/2)(C_N+2T-1)-1$ | $T_A + [1+\log_2(C_N-k+2T-1)]T_X$ |

Note that the proposed method may not give a better space complexity in the case where $k < 2T+1$, e.g. type 3 GNB for $GF(2^4)$. Nonetheless, $k$ is typically greater than $2T$.

## VI. CONCLUSION

This paper presents the matrix decomposition method for odd-type Gaussian normal basis multiplication for $GF(2^k)$. The property of odd-type Gaussian normal basis is proved, and the matrix decomposition method for optimal normal basis is adapted for odd-type Gaussian normal basis. The proposed method is compared with previous works. Our implementation uses less number of XOR gates with a small trade-off in the critical path delay. We plan to construct a method to reduce the space complexity of odd-type Gaussian normal basis multiplier further without increasing the critical path delay, and apply the matrix decomposition method to even-type Gaussian normal basis multiplier, which is already suggested by Kizilkale et al. in 2016, as our future works.

## ACKNOWLEDGMENT

This research was supported by Chula Computer Engineering Graduate Scholarship for CP Alumni, Department of Computer Engineering, Faculty of Engineering, Chulalongkorn University.

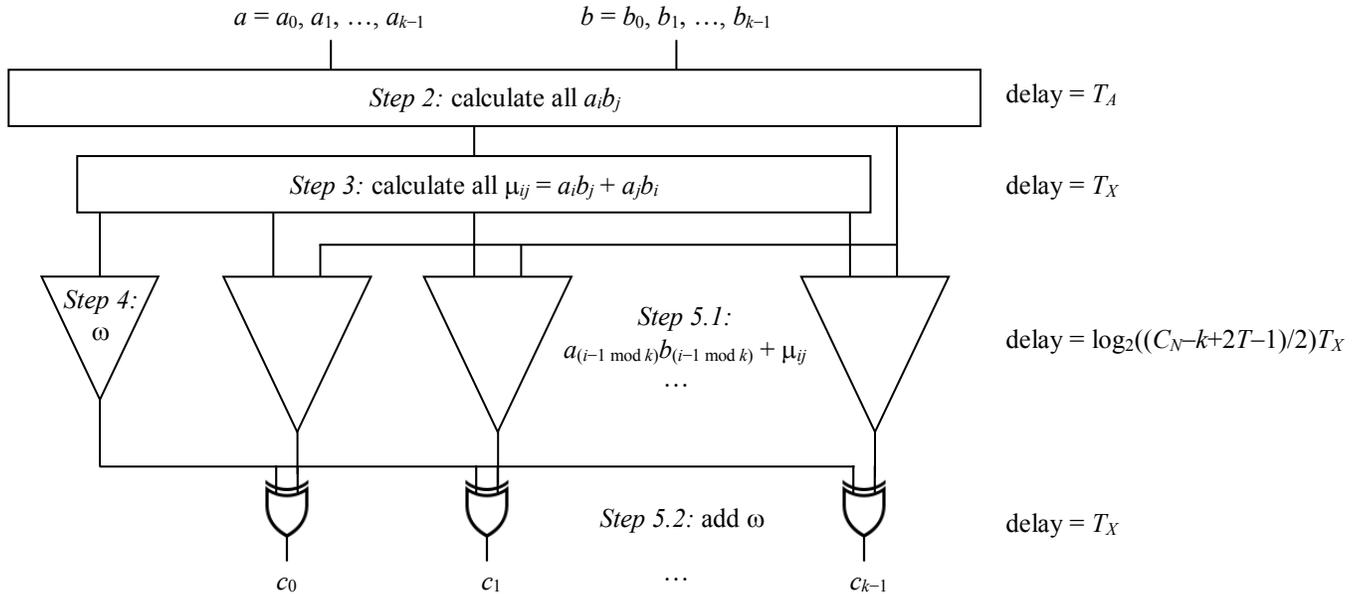

Figure 1. The design of odd-type GNB multiplier for $GF(2^k)$ using our proposed decomposition method.